\documentclass[12pt]{amsart}
\usepackage{geometry}
\usepackage{setspace}
\usepackage{times}
\usepackage{fullpage}
\usepackage{graphicx}
\usepackage{epstopdf}
\makeatletter
\def\specialsection{\@startsection{section}{1}%
  \z@{\linespacing\@plus\linespacing}{.5\linespacing}%
  {\normalfont}}
\def\section{\@startsection{section}{1}%
  \z@{.7\linespacing\@plus\linespacing}{.5\linespacing}%
  {\normalfont\scshape}}
\makeatother

\title{O\MakeLowercase{xypnictide} S\MakeLowercase{m}F\MakeLowercase{e}A\MakeLowercase{s}(O,F) \MakeLowercase{superconductor: a candidate for high--field magnet applications}}
\author{Kazumasa Iida$^{1,*}$ \and Jens H\"{a}nisch$^1$ \and Chiara Tarantini$^2$ \and Fritz Kurth$^1$ \and Jan Jaroszynski$^2$ \and Shinya Ueda$^3$ \and Michio Naito$^3$ \and Ataru Ichinose$^4$ \and Ichiro Tsukada$^4$ \and Elke Reich$^1$ \and Vadim Grinenko$^1$ \and Ludwig Schultz$^1$ \and Bernhard Holzapfel$^1$}
\begin{document}
\maketitle
\doublespacing

{\footnotesize
\noindent
1. Institute for Metallic Materials, IFW Dresden, 01171 Dresden, Germany\\
2. Applied Superconductivity Center, National High Magnetic Field Laboratory, Florida State University, 2031 East Paul Dirac Drive, Tallahassee, Florida 32310, USA\\
3. Department of Applied Physics, Tokyo University of Agriculture and Technology, Koganei, Tokyo 184-8588, Japan\\
4. Central Research Institute of Electric Power Industry, 2-6-1 Nagasaka, Yokosuka, Kanagawa 240-0196, Japan\\
$^{*}$Corresponding author. k.iida@ifw-dresden.de} 

\section*{Abstract}
\noindent{\bf The recently discovered oxypnictide superconductor SmFeAs(O,F) is the most attractive material among the Fe--based superconductors due to its highest transition temperature of 56\,K and potential for high-field performance. In order to exploit this new material for superconducting applications, the knowledge and understanding of its electro-magnetic properties are needed. Recent success in fabricating epitaxial SmFeAs(O,F) thin films opens a great opportunity to explore their transport properties. Here we report on a high critical current density of over 10$^5$\,A/cm$^2$ at 45\,T and 4.2\,K for both main field orientations, feature favourable for high-field magnet applications. Additionally, by investigating the pinning properties, we observed a dimensional crossover between the superconducting coherence length and the FeAs interlayer distance at 30--40\,K, indicative of a possible intrinsic Josephson junction in SmFeAs(O,F) at low temperatures that can be  employed in electronics applications such as a terahertz radiation source and a superconducting Qubit.}

\section*{Introduction}
Among the recently discovered Fe-based superconductors\cite{01}, the highest superconducting transition temperature $T_{\rm c}$ of 56\,K has been reported in SmFeAs(O,F)\cite{02}. This new class of material shows very high upper critical fields at low temperatures together with a moderate anisotropy ranging from 4 to 7\cite{Lee}, which is suitable for high-field magnet applications. Hence several attempts on wire fabrication using SmFeAs(O,F) by powder-in-tube technique (PIT) have already been reported\cite{03}, despite the lack of information on the field and orientation dependence of intra-grain critical current density [i.e., $J_{\rm c}(H,\Theta)$]. In order to exploit this material class, the knowledge of these properties should be clarified.

Epitaxial thin films are favourable for electronics device applications and investigating transport as well as optical properties thanks to their geometry. Recent success in fabricating epitaxial Fe-based superconducting thin films opens a great opportunity for investigating their physical properties and exploring possible superconducting applications. To date, high-field transport properties of Co-doped SrFe$_2$As$_2$ (Sr-122) and BaFe$_2$As$_2$ (Ba-122), and Fe(Se,Te) epitaxial thin films have been reported by several groups\cite{Baily,Jens-1,Tarantini-1}. For Co-doped Ba-122, $J_{\rm c}$ performance can be tuned by introduction of artificial pinning centers and proton irradiation\cite{04,Boris}. Additionally, multilayer approaches that can tailor superconducting properties and their anisotropy have been reported by Lee $et$ $al$\cite{Lee-2}. Furthermore, epitaxial Co-doped Ba-122 and Fe(Se,Te) thin films have been realised on ion beam assisted deposition MgO coated conductor templates\cite{05,06,07} and the rolling-assisted biaxially textured substrate\cite{08}, respectively. Similarly, high performance K-doped Ba-122 and Sr-122 wires by PIT  have been reported by Weiss $et$ $al$.\cite{Weiss} and Gao $et$ $al$\cite{Gao}, respectively. These results are very promising for realising Fe-based superconducting high-field applications. However, transport critical current properties of high-$T_{\rm c}$ (i.e., over 50\,K) oxypnictide thin films have not been reported before due to the absence of high quality films. Recently, $in$ $situ$ prepared $Ln$FeAs(O,F) ($Ln$=Nd and Sm) epitaxial thin films with $T_{\rm c}$ exceeding 50\,K have been realised by molecular beam epitaxy (MBE)\cite{10,Uemura}. These successes give many possibilities to explore electro-magnetic properties.

In this paper, we report on various $in-plane$ (i.e., current is flowing on the crystallographical $ab$-plane) transport properties up to 45\,T of epitaxial SmFeAs(O,F) thin films grown by MBE on CaF$_2$ (001) single crystalline substrates and discuss their pinning properties. A high $J_{\rm c}$ of over 10$^5$\,A/cm$^2$ was recorded at 45\,T and 4.2\,K for both crystallographic directions, which is favourable for high-field magnet applications. By analysing pinning properties the dimensional crossover between the out-of-plane superconducting coherence length $\xi_{\rm c}$ and the Fe-As interlayer distance $d_{\rm FeAs}$ was observed at 30-40\,K. This indicates the possible intrinsic Josephson junction in SmFeAs(O,F) at low temperatures.

\section*{Results}
\subsection*{Microstructural analyses}
As verified by x-ray diffraction, the biaxially textured SmFeAs(O,F) film with a narrow full width at half maximum (FWHM) of less than 0.65$^\circ$ was obtained (See in Supplementary Fig.\,S1). As shown in Fig.\,\ref{fig:figure1}a, trapezoid shaped Sm(O,F) cap layers, which are changed from SmF$_3$, are aligned discontinuously. Additionally, a crystallographically disordered layer with around 20\,nm thickness as indicated by the arrows is present between Sm(O,F) cap and SmFeAs(O,F) layers. Relatively dark particles are observed in the SmFeAs(O,F) matrix, which are identified as iron-fluoride, presumably FeF$_2$, by elemental mappings shown in Figs.\,\ref{fig:figure1}b and \ref{fig:figure1}c. This is due to the excess of Fe supplied during the film growth.

Compared to Fe(Se,Te)\cite{Tsukada} and Co-doped Ba-122 films\cite{Fritz} grown by pulsed laser deposition, a relatively sharp and clean interface is observed between SmFeAs(O,F) and CaF$_2$ substrate, as shown in Fig.\,\ref{fig:figure1}d. Furthermore, SmFeAs(O,F) layers contained neither correlated defects nor large angle grain boundaries (GBs).

\subsection*{Resistivity measurements up to 45\,T}
The superconducting transition temperature $T_{\rm c,90}$ defined as 90\,\% of the normal state resistivity is 54.2\,K in zero magnetic field. Figures\,\ref{fig:figure2}a and \ref{fig:figure2}b show the Arrhenius plots of resistivity for both crystallographic directions measured in static fields up to 45\,T. For both directions the $T_{\rm c,90}$ is shifted to lower temperature with increasing $H$, as shown in the inset of Fig.\,\ref{fig:figure2}b. The respective $T_{\rm c,90}$ at 45\,T for $H\parallel c$ and $\parallel ab$ are 44.9\,K and 49.9\,K. Significant broadening of the transition is observed for $H\parallel c$, which is reminiscent of high-$T_{\rm c}$ cuprates. Such broadening of the transition originates from enhanced thermally activated vortex motion for $H\parallel c$. In contrast, the in-field $T_{\rm c,90}$ as well as its transition width for $H\parallel ab$ are less affected by $H$ than that for $H\parallel c$.

The activation energy $U_0(H)$ for vortex motion can be estimated by the model of thermally activated flux flow\cite{13}. On the assumption that $U(T,H)=U_0(H)(1-T/T_{\rm c})$, we obtain ${\rm ln}\rho(T,H)= {\rm ln}\rho_0(H)-U_0(H)/T$ and ${\rm ln}\rho_0(H)={\rm ln}\rho_{\rm 0f}+U_0(H)/T_{\rm c}$, where  $\rho_{\rm 0f}$ is the prefactor. In Figs.\,\ref{fig:figure2}a and \ref{fig:figure2}b, the slope of linear fits corresponds to the $U_0$ for vortex motion. Figure\,\ref{fig:figure2}c shows $U_0$ as a function of $H$ for both major directions. It can be seen that $U_0(H)$ shows a power law [i.e., $U_0(H)\sim H^{-\alpha}$] for both crystallographic directions. In the range of  $1<\mu_{0}H<8$\,T, $\alpha=0.46$ is observed for $H\parallel c$, whilst a similar field dependence of $U_0(H)$ reaches 20\,T for $H\parallel ab$. In higher fields $U_0$ for $H\parallel ab$ shows a weak $H$ dependence. On the other hand, $\alpha=1.2$ is obtained for $H\parallel c$ in the range of $8<\mu_{0}H<45$\,T, which is close to 1, suggesting a crossover from plastic to collective pinning at around $\mu_{0}H\sim8$\,T\cite{Yeshurun}.

Figure\,\ref{fig:figure2}d shows the relationship between ${\rm ln}\rho_0$ and $U_0$ for $H\parallel c$ and $\parallel ab$. The linear fitting for $H\parallel c$ yields $T_{\rm c}=53.4\pm0.2$\,K, whilst the corresponding value for $H\parallel ab$ is $T_{\rm c}=53.5\pm0.2$\,K. Both $T_{\rm c}$ values are equal within error and close to $T_{\rm c,90}$. This perfect linear scaling is due to the correct assumption that both $U(T,H)=U_0(H)(1-T/T_{\rm c})$ and $\rho_{\rm 0f}={\rm const.}$ conditions are satisfied in a wide temperature range in Figs.\,\ref{fig:figure2}a and \ref{fig:figure2}b.

\subsection*{In-field $J_{\rm c}$ performance}
The field dependence of $J_{\rm c}$ at 4.2\,K for both principal crystallographic directions measured up to 45\,T is displayed in Fig.\,\ref{fig:figure3}a. $J_{\rm c}$ for $H\parallel c$ ($J^{c}_{\rm c}$) is lower than that for $H\parallel ab$ ($J^{ab}_{\rm c}$), which is a consequence of moderate anisotropy of SmFeAs(O,F). This tendency is observed for all temperature regions (see Supplementary Fig.\,S2). It is worth mentioning that a $J^{c}_{\rm c}$ of over 10$^5$\,A/cm$^2$ was recorded even at 45\,T, which is favourable for high-field magnet applications.

$J^{ab}_{\rm c}$ is observed to decrease gradually with $H$ and it shows an almost constant value of $7.4\times10^5$\,A/cm$^2$ for $\mu_{0}H>28$\,T. This behaviour can be explained by a combination of extrinsic (i.e., normal precipitates and stacking faults) and intrinsic pinning, which is a similar observation in quasi two-dimensional (2D) system YBa$_2$Cu$_3$O$_{7-\delta}$ [i.e., $\xi_{\rm c}(0)/d_{\rm CuO_2}\sim 0.4$, where $\xi_{\rm c}(0)$ is the out-of-plane superconducting coherence length at zero temperature and $d_{\rm CuO_2}$ is the interlayer distance between CuO$_2$ planes]\cite{Awaji}. SmFeAs(O,F) is an alternating structure of SmO and FeAs layers, similarly to high-$T_{\rm c}$ cuprates. Additionally, $\xi_{\rm c}(0)$ is shorter than the interlayer distance between Fe-As planes $d_{\rm FeAs}$. Hence, modulation of superconducting order parameter along the crystallographic $c$-axis (i.e., intrinsic pinning) is highly expected in SmFeAs(O,F). In fact the extrinsic pinning is dominant up to 28\,T, whereas the intrinsic pinning overcomes the extrinsic one above 28\,T. The estimation of $\xi_{\rm c}(0)$ and $d_{\rm FeAs}$ in our SmFeAs(O,F) case will be discussed later.

By analysing the $E$-$J$ curves from which $J_{\rm c}$ was determined, we obtain the information on the pinning potential. On the assumption of a logarithmic current dependence of the pinning potential $U_{\rm p}$ for homogeneous samples, $E$-$J$ curves show a power-law relation $E$$\sim$$J^n$ ($n\sim U_{\rm p}/k_{\rm B}T$, where $k_{\rm B}$ is the Boltzmann constant)\cite{Zeldov}. Hence $J_{\rm c}$ scales with $n$ and indeed the field dependence of $n$ has a similar behaviour to $J_{\rm c}(H)$ for $H\parallel c$, as presented in Fig.\,\ref{fig:figure3}b. For $H\parallel ab$, $n$ decreases with $H$ up to 28\,T, similarly to the $J_{\rm c}(H)$ behaviour, whereas at larger field it suddenly increases due to the dominating intrinsic pinning. Hence a failure to scale $J_{\rm c}$ with $n$ or deviations as shown in Fig.\,\ref{fig:figure3}c indicates the presence of intrinsic pinning. 

The field dependence of the pinning force density $F_{\rm p}$ for both crystallographic directions at 4.2\,K is summarised in Fig.\,\ref{fig:figure3}d. An almost field independent $F_{\rm p}$ above 10\,T for $H\parallel c$ is observed, whereas $F_{\rm p}$ for $H\parallel ab$ is still increasing up to the maximum field available.

\subsection*{Angular dependence of $J_{\rm c}$}
In order to gain a deeper insight into the flux pinning, the angular dependence of $J_{\rm c}$ [$J_{\rm c}(\Theta)$, where $\Theta$ is the angle between $H$ and the $c$-axis] was measured and summarised in Fig.\,\ref{fig:figure4}. Figure\,\ref{fig:figure4}a presents $J_{\rm c}(\Theta)$ at 30\,K in three different magnetic fields. Almost isotropic $J_{\rm c}(\Theta,2.5{\rm T})$ of around 0.14 MA/cm$^2$ was observed at angles $\Theta$ up to $75^\circ$. Similar isotropic behaviour is seen at 6\,T. These results suggest the presence of $c$-axis correlated defects. However, the presence of these defects is ruled out by TEM investigation, since only relatively large FeF$_2$ particles are observed in the SmFeAs(O,F) matrix. Recently, van der Beek $et$ $al$. pointed out that defects of size larger than the out-of-plane coherence length contribute to $c$-axis pinning in anisotropic superconductors\cite{Beek}. Additionally, the intrinsic pinning is active below $T=30\sim40$\,K, as shown below. Hence the combination of large particles and the intrinsic pinning may be responsible for this isotropic $J_{\rm c}(\Theta)$.  

For $H\parallel ab$, a broad maximum of $J_{\rm c}$ is observed and this peak becomes sharper with increasing $H$ (Fig.\,\ref{fig:figure4}a). However, the corresponding $n$ shows a broad minimum for $H$ close to $ab$ direction (Fig.\,\ref{fig:figure4}b), which is opposite behaviour to $J_{\rm c}$. This is due to the thermal fluctuation of Josephson vortices, which leads to flux creep. Here, the flux creep rate $S=-dln(J)/dln(t)$ and the exponent $n$ are related as $S=1/(n-1)$\cite{Yamasaki}. When the applied field is close to the $ab$-plane, a number of thermally fluctuated Josephson vortices are generated, leading to an increase in $S$. This could quantitatively explain a dip of $n$ at around $H$ close to $ab$. Similar behaviour has been observed in YBa$_2$Cu$_3$O$_{7-\delta}$ thin films\cite{18,Awaji-2,19} and Fe(Se,Te) thin films\cite{20}. On the other hand, this dip of $n$ disappears at 40\,K, although the $J_{\rm c}$ still shows a broad maximum (Figs.\,\ref{fig:figure4}c and \ref{fig:figure4}d). Hence the activation temperature of the intrinsic pinning is between 30 and 40\,K, which is in good agreement with the transition temperature between Abrikosov- and Josephson-like vortices in SmFeAs(O,F) single crystals\cite{21}.

Figure\,\ref{fig:figure4}e shows $J_{\rm c}(\Theta)$ measured at 4.2\,K in fields up to 40\,T. A sharp peak is observed for $H\parallel ab$ with a $J_{\rm c}$ of around of $8\times10^5$\,A/cm$^2$. For 2D superconductors (e.g., Bi$_2$Sr$_2$CaCu$_2$O$_{8+x}$), the relation $J_{\rm c}(\Theta, H)=J^{c}_{\rm c}(H{\rm cos}\Theta)$ holds in the intrinsic pinning regime, whereas $J^{ab}_{\rm c}$ is field independent\cite{Tachiki,Schmitt}. Thus, in this regime $J_{\rm c}(\Theta, H)$ depends only on the field component along the $c$-axis. For our SmFeAs(O,F) thin film, the aforementioned condition is satisfied above 28\,T at which the crossover field between extrinsic and intrinsic pinning is observed (see Fig.\,\ref{fig:figure3}a). Hence, for $\Theta> 59^\circ$ (${\rm sin}\Theta=28/32.5=0.86$, $\Theta={\rm sin}^{-1}(0.86)=59^\circ$) the $ab$ component of the applied fields exceed 28\,T, entering in the field-independent $J^{ab}_{\rm c}$ region. It means that both angular--$J_{\rm c}$ curves measured at 32.5 and 40\,T rescale with $H{\rm cos}\Theta$, as shown in Fig.\,\ref{fig:figure4}f.

\section*{Discussion}
We estimate the $\xi_{\rm c}(0)$ by using $T_{\rm cr}=(1-\tau_{\rm cr})T_{\rm c}$, where $T_{\rm cr}$ is the dimensional crossover temperature and $\tau_{\rm cr}=2\xi_{\rm c}(0)^2/d_{\rm FeAs}^2$ is the dimensionless ratio characterising the crossover from quasi-2D layered to continuous 3D anisotropic behaviour\cite{Blatter}. By substituting $T_{\rm cr}=30-40$\,K and $d_{\rm FeAs}=0.858$\,nm from the x-ray diffraction shown in Supplementary Fig.\,S1, $\xi_{\rm c}(0)= d_{\rm FeAs}\sqrt{(1-\frac{T_{\rm cr}}{T_{\rm c}})/2}$ is calculated to 0.3$\sim$0.4\,nm. The ratio $\xi_{\rm c}(0)/d_{\rm FeAs}=0.35\sim0.47$ explains the intrinsic pinning related to a quasi 2D system observed in this film. The relation $\xi_{\rm ab}(0)\simeq\sqrt\gamma\xi_{\rm c}(0)$ yields $\xi_{\rm ab}(0)=1.7\sim2.2$\,nm, where $\gamma$ is the effective-mass or resistivity anisotropy, which is about 30 at $T=0$\,K from measurements of the $c$-axis plasma frequency using infrared ellipsometry\cite{Dubroka}. The evaluated superconducting coherence lengths for both crystallographic directions are in very good agreement with single-crystals values reported by Welp $et$ $al$\cite{welp}.

The presence of a dimensional crossover indicates a possible intrinsic Josephson junction in SmFeAs(O,F), which can be used in superconducting electronics applications such as a terahertz radiation source and a superconducting Qubit\cite{Ozyuzer, Kubo}. Indeed, the intrinsic Josephson junction was reported for a PrFeAsO$_{0.7}$ single crystal, where an $s$-shaped stack junction in $c$-direction was prepared by focused ion beam\cite{Kashiwaya}.

For high-field magnet applications, a high $J_{\rm c}$ together with a low $J_{\rm c}$ anisotropy ($\frac{J^{ab}_{\rm c}}{J^{c}_{\rm c}}$) in the presence of magnetic field is necessary. The present results are promising, since $J_{\rm c}$ is over 10$^5$\,A/cm$^2$ at 45\,T for both crystallographic directions. Further increasing in $J_{\rm c}$ is possible, since the only appreciable defects in our SmFeAs(O,F) films are large FeF$_2$ particles. Improved pinning performance and, as a consequence, larger $J_{\rm c}$ could be realised by incorporating artificial pinning centres similarly to Co-doped Ba-122 thin films reported by Tarantini $et$ $al$\cite{04}. Albeit the $J_{\rm c}$ anisotropy is increasing with $H$, this value is still low compared to high-$T_{\rm c}$ cuprtaes. For instance, $J_{\rm c}$ anisotropy is about 3.6 at 30\,T and 4.2\,K in SmFeAs(O,F), whereas the corresponding value in YBa$_2$Cu$_3$O$_{7-\delta}$ is over 7, albeit the latter shows higher $J_{\rm c}$ than the former\cite{Xu}.

PIT is a more realistic process than MBE for high-field magnet applications. High temperature heat treatment in PIT leads to a loss of F, however, this problem can be solved by employing a low temperature synthesis and ex-situ process with SmF$_3$ containing binder as explained in refs.\cite{Ma,Fujioka}. Despite a high $T_{\rm c}$ of over 45\,K for both SmFeAs(O,F) wires, self-field $J_{\rm c}$ shows only a few thousand A/cm$^2$ at 4.2\,K, which is presumably due to grain boundaries (GBs), poor grain connectivity and low density. Obviously these PIT processed wires contain a high density of large angle GBs. In the case of Co-doped Ba-122 GBs with misorientation angles above 9$^\circ$ seriously reduce the critical current. \cite{Katase}. However, PIT processed K-doped Ba-122 and Sr-122 wires showed a relatively high inter-grain $J_{\rm c}$\cite{Weiss, Gao}. Clean GBs (i.e., no segregation of secondary phases around GBs), good grain connectivity and a low anisotropy may be responsible for these high performance wires. An approach similar to the one employed in K-doped Ba-122 and Sr-122 wires fabrication may be useful for improving inter-grain $J_{\rm c}$ in SmFeAs(O,F) wires as well. Nevertheless bicrystal experiments on SmFeAs(O,F) will give a valuable information on these issues.

To conclude, we have explored intrinsic electro-magnetic properties of epitaxial SmFeAs(O,F) thin films prepared by MBE on CaF$_2$ (001) substrate by measuring field-angular dependence of transport properties up to 45\,T. Our findings strongly support the presence of a competition behaviour between extrinsic pinning below 28\,T and intrinsic pinning above 28\,T. We also determined that the intrinsic pinning starts being effective below $T=30\sim40$\,K, at which the crossover between the out-of-plane coherence length and the interlayer distance occurs. This knowledge of SmFeAs(O,F) electro-magnetic properties could stimulate future development of superconducting applications of this class of material.

\section*{Methods}
\subsection*{Epitaxial SmFeAs(O,F) film preparation by MBE}
SmFeAs(O,F) films of 80\,nm thickness have been grown in the customer-designed MBE chamber. A parent compound of SmFeAsO film was prepared on CaF$_2$ (001) single crystalline substrate at 650\,$^\circ$C, followed by the deposition of a SmF$_3$ cap layer. Empirically, Fe-rich pnictide films fabricated by MBE showed high $J_{\rm c}$ values\cite{Ikuta}. Hence a slight Fe excess was supplied during the growth of SmFeAsO layers. After the overlayer deposition, the sample was kept at the same temperature in the MBE chamber for 0.5 h for the purpose of F diffusion into the SmFeAsO layer. The detailed fabrication process can be found in ref.\cite{10}. SmFeAs(O,F) films are grown epitaxially with high crystalline quality confirmed by x-ray diffraction, which is summarised in Supplementary Fig.\,S1.

\subsection*{Microstructural analyses by TEM}
A TEM lamella was prepared by means of focused ion beam. Microstructural analyses have been performed by using a JEOL TEM-2100F transmission electron microscope equipped with an energy-dispersive x-ray spectrometer.

\subsection*{In-plane transport properties measurement}
A small bridge of 70\,$\mu$m width and 0.7\,mm length was fabricated by laser cutting. $I$-$V$ characteristics on this sample were measured with four-probe configuration by a commercial physical property measurement system [(PPMS) Quantum Design] up to 12\,T. Transport measurements up to 45\,T were carried out in the high field dc facility at the National High Magnetic Field Laboratory (NHMFL) in Tallahassee, FL. A voltage criterion of 1\,$\rm\mu V/cm$ was employed for evaluating $J_{\rm c}$. The magnetic field $H$ was applied in maximum Lorentz force configuration during all measurements ($H \perp J$, where $J$ is current density).

\subsection*{Acknowledgement} The authors would like to thank M.\,Weigand and B.\,Maiorov of Los Alamos National Laboratory, D.\,C.\,Larbalestier of Applied Superconductivity Center, National High Magnetic Field Laboratory, Florida State University for fruitful discussions and comments, as well as M.\,K\"{u}hnel and U.\,Besold for their technical support. The research leading to these results has received funding from European Union's Seventh Framework Programme (FP7/2007-2013) under grant agreement number 283141 (IRON-SEA). A portion of this work was performed at the National High Magnetic Field Laboratory, which is supported by National Science Foundation Cooperative Agreement No. DMR-0654118, the State of Florida, and the U.S. Department of Energy. This research has been also supported by Strategic International Collaborative Research Program (SICORP), Japan Science and Technology Agency. V.G. acknowledges financial support of the EU (Super Iron under project No. FP7-283204).

\subsection*{Authors contribution} K.I., J.H. and C.T. designed the study and wrote the manuscript together with M.N., J.J., I.T., V.G., L.S., and B.H. Thin films were prepared by S.U. K.I. and S.U. conducted x-ray experiments. C.T. and K.I. measured low field transport properties. C.T., K.I., J.H., F.K., M.N., and J.J. investigated high-field transport properties. A.I., I.T., and E.R. conducted TEM investigation. All authors discussed the results and implications and commented on the manuscript.

\subsection*{Additional information}
The authors declare no competing financial interests. Correspondence and requests for materials should be addressed to K. I.

\newpage
\section*{Figure legends}

\subsection*{Figure 1 Microstructural analyses by TEM} (a) Cross-sectional scanning TEM image of the SmFeAs(O,F) thin film. A crystallographically disordered layer as indicated by the arrows is present between Sm(O,F) cap and SmFeAs(O,F) layers. (b) Elemental Fe and (c) F mappings measured by energy dispersive x-ray spectroscopy. (d) High-resolution TEM image of the SmFeAs(O,F) thin film in the vicinity of the CaF$_2$ substrate / SmFeAs(O,F) film interface.

\subsection*{Figure 2 In-field resistivity ($\rho$) measurements of SmFeAs(O,F) film up to 45\,T and the analyses of the activation energy of pinning potential ($U_0$)} (a) Arrhenius plots of $\rho$ at various magnetic fields parallel to the crystallographic $c$-axis and (b) $ab$-plane. The inset shows the $\mu_{0}H-T_{\rm c,90}$ diagram of SmFeAs(O,F) film for both directions, which is the identical to the extracted temperature dependence of the upper critical fields by employing a 90\,\% criterion of the normal state resistivity. (c) Field dependence of the activation energy for $H\parallel c$ and $\parallel ab$. (d) Relationship between ${\rm ln}\rho_0$ and $U_0$ for $H\parallel c$ and $\parallel ab$.

\subsection*{Figure 3 In-field critical current density ($J_{\rm c}$) performance of SmFeAs(O,F) thin film at 4.2\,K} (a) Field dependence of $J_{\rm c}$ measured at 4.2\,K up to 45\,T for both crystallographic directions and (b) the corresponding exponent $n$ values. A crossover from extrinsic to intrinsic pinning is shown by the arrow. (c) Scaling behaviour of the field dependent $J_{\rm c}$. (d) The pinning force density $F_{\rm p}$ for both crystallographic directions at 4.2\,K.

\subsection*{Figure 4 Field and orientation dependence of critical current density ($J_{\rm c}$) of SmFeAs(O,F) thin film} (a) Angular dependence of $J_{\rm c}$ measured at 3 different applied magnetic fields at 30\,K and (b) the corresponding exponent $n$ values. (c) $J_{\rm c}(\Theta,H)$ measured at 40\,K under several magnetic fields ($\mu_{0}H=2.5, 4$ and 6\,T) and (d) the corresponding exponent $n$ values. (e) Angular dependence of $J_{\rm c}$ at 4.2\,K under various applied magnetic fields up to 40\,T. (f) Scaling behaviour of the angular dependent $J_{\rm c}$ measurements. Below 17\,T (i.e., by substituting $\mu_0H=32.5$\,T and $\Theta=59^\circ$ in $H{\rm cos}\Theta$) as indicated by the arrow, both curves overlap each other.

\newpage

\begin{figure}
	\centering
		\includegraphics[width=13cm]{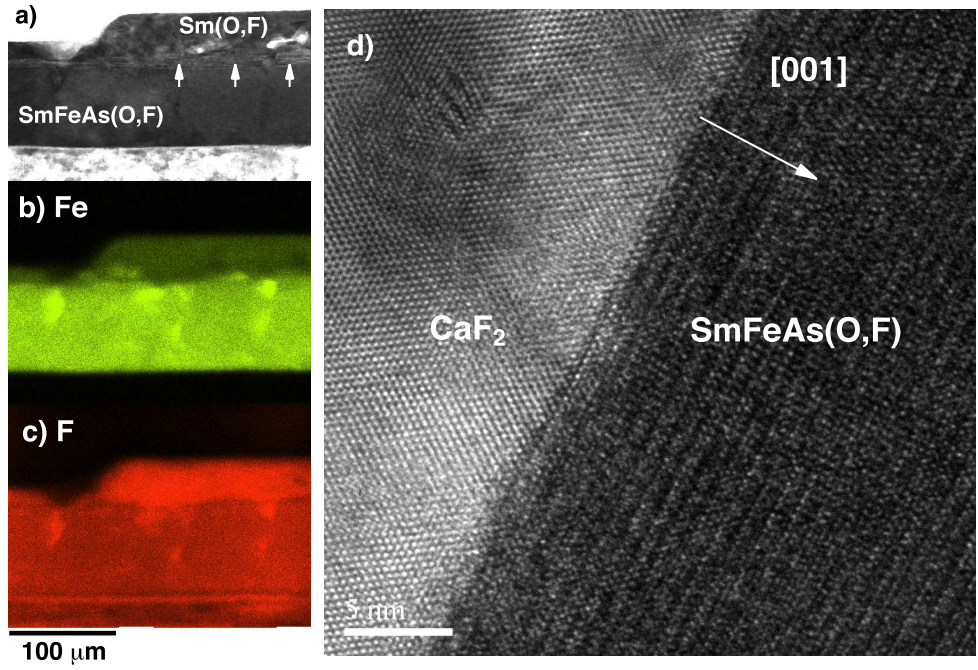}
		\caption{Iida $et$ $al$.}
\label{fig:figure1}
\end{figure}

\clearpage
\begin{figure}
	\centering
		\includegraphics[width=13cm]{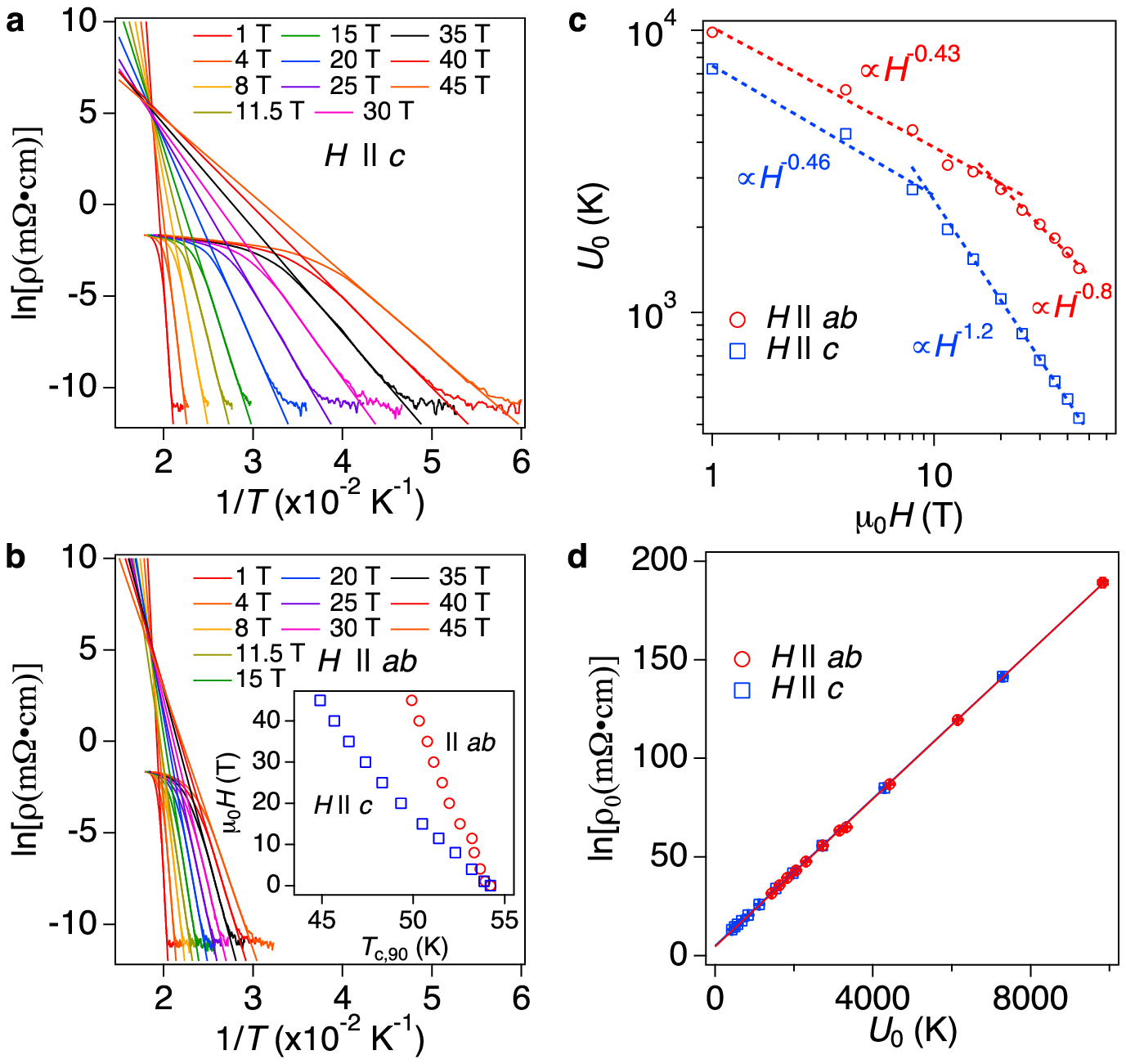}
		\caption{Iida $et$ $al$.}
\label{fig:figure2}
\end{figure}

\clearpage
\begin{figure}
	\centering
		\includegraphics[width=13cm]{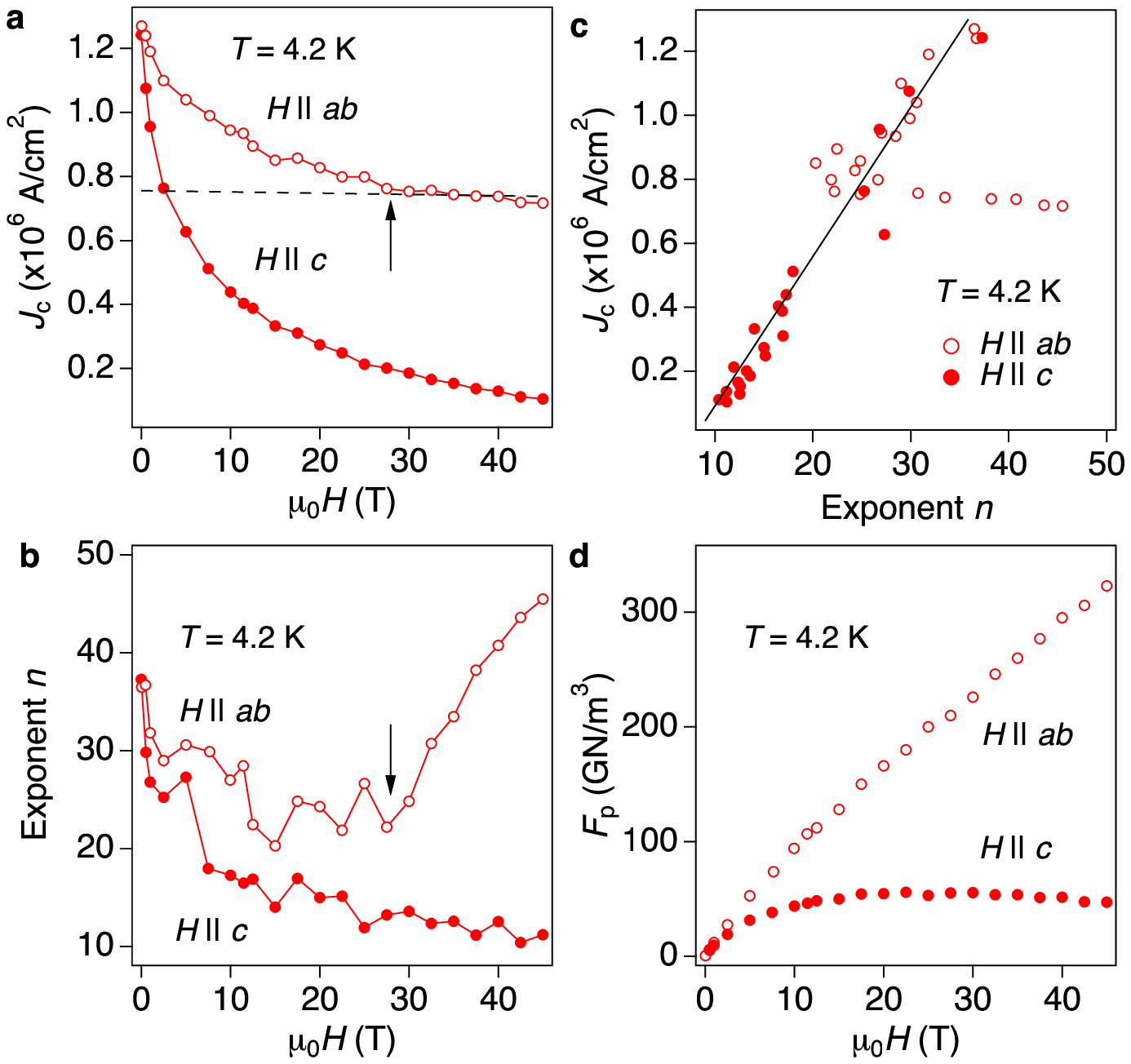}
		\caption{Iida $et$ $al$.} 
\label{fig:figure3}
\end{figure}

\clearpage
\begin{figure}
	\centering
			\includegraphics[width=13cm]{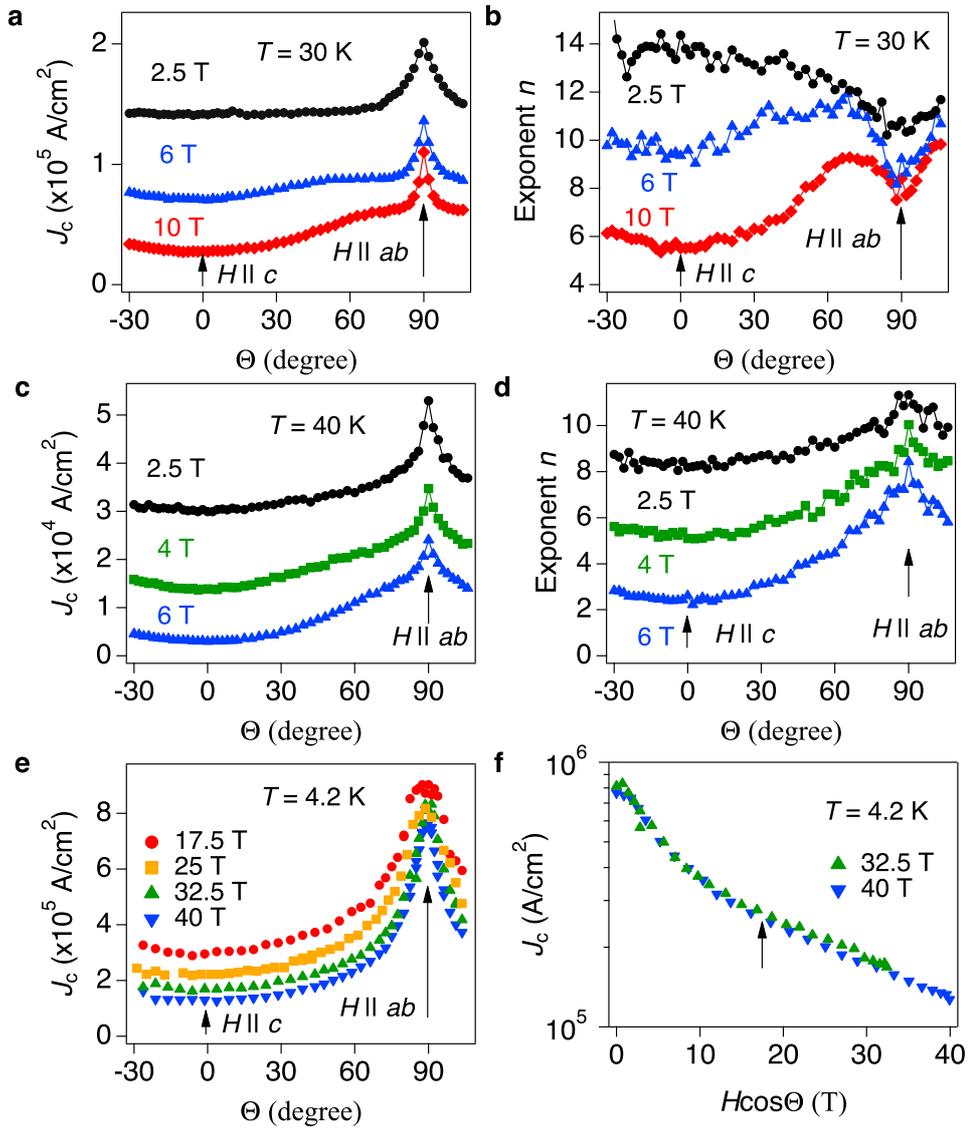}
		\caption{Iida $et$ $al$.} 
\label{fig:figure4}
\end{figure}

\clearpage

\renewcommand{\thefigure}{S\arabic{figure}}
\setcounter{figure}{0}
\subsection*{Supplementary information: Oxypnictide SmFeAs(O,F) superconductor: a candidate for high-field magnet applications}
\begin{figure}[b]
	\centering
			\includegraphics[width=10cm]{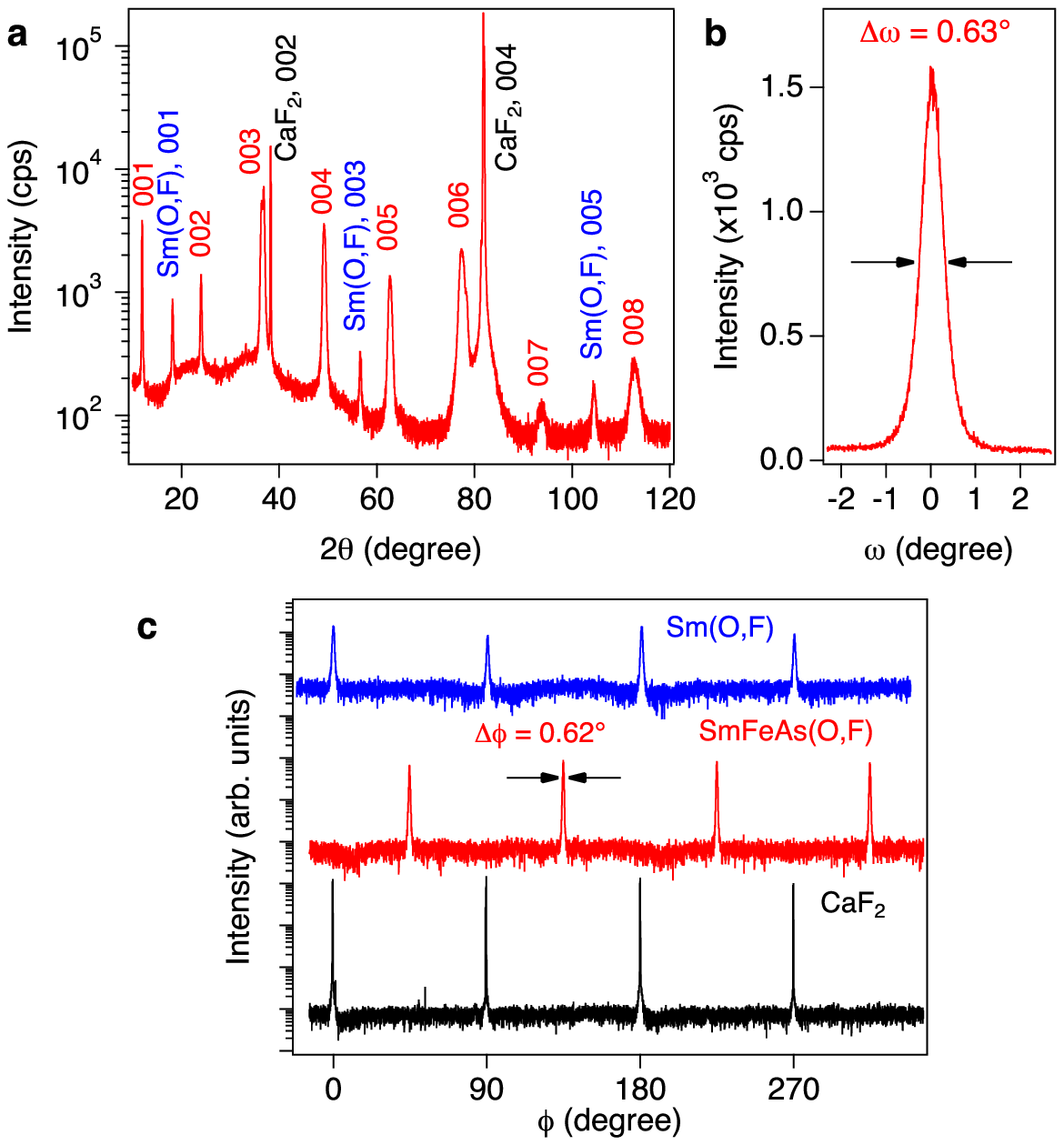}
		\caption{(a) The $\theta/2\theta$ scan of the SmFeAs(O,F) thin film grown on CaF$_2$ (001) substrate. All peaks are assigned as the 00$l$ reflections of SmFeAs(O,F), Sm(O,F) and CaF$_2$, indicative of the $c$-axis textured for both SmFeAs(O,F) and Sm(O,F) layers. (b) The rocking curve of the 004 reflection of SmFeAs(O,F) shows a narrow full width at half maximum (FWHM) of 0.63$^\circ$, proving a highly out-of-plane textured film. (c) $\phi$ scans of the 112 peak of SmFeAs(O,F), the 112 peak of Sm(O,F), and the 111 peak of CaF$_2$. The $\phi$ scan of SmFeAs(O,F) revealed no additional reflections other than sharp (average FWHM, $\Delta\phi$, of 0.63$^\circ$) and strong reflections at every 90$^\circ$, indicative of biaxial texture. These results highlight that the SmFeAs(O,F) is grown epitaxially with high crystalline quality. Interestingly Sm(O,F) cap layer is also grown biaxial textured. The epitaxial relation of each layer and substrate is confirmed to (001)[100]Sm(O,F)$\|$(001)[110]SmFeAs(O,F)$\|$(001)[100]CaF$_2$.} 
\label{fig:SA1}
\end{figure}

\begin{figure}
	\centering
		\includegraphics[width=13cm]{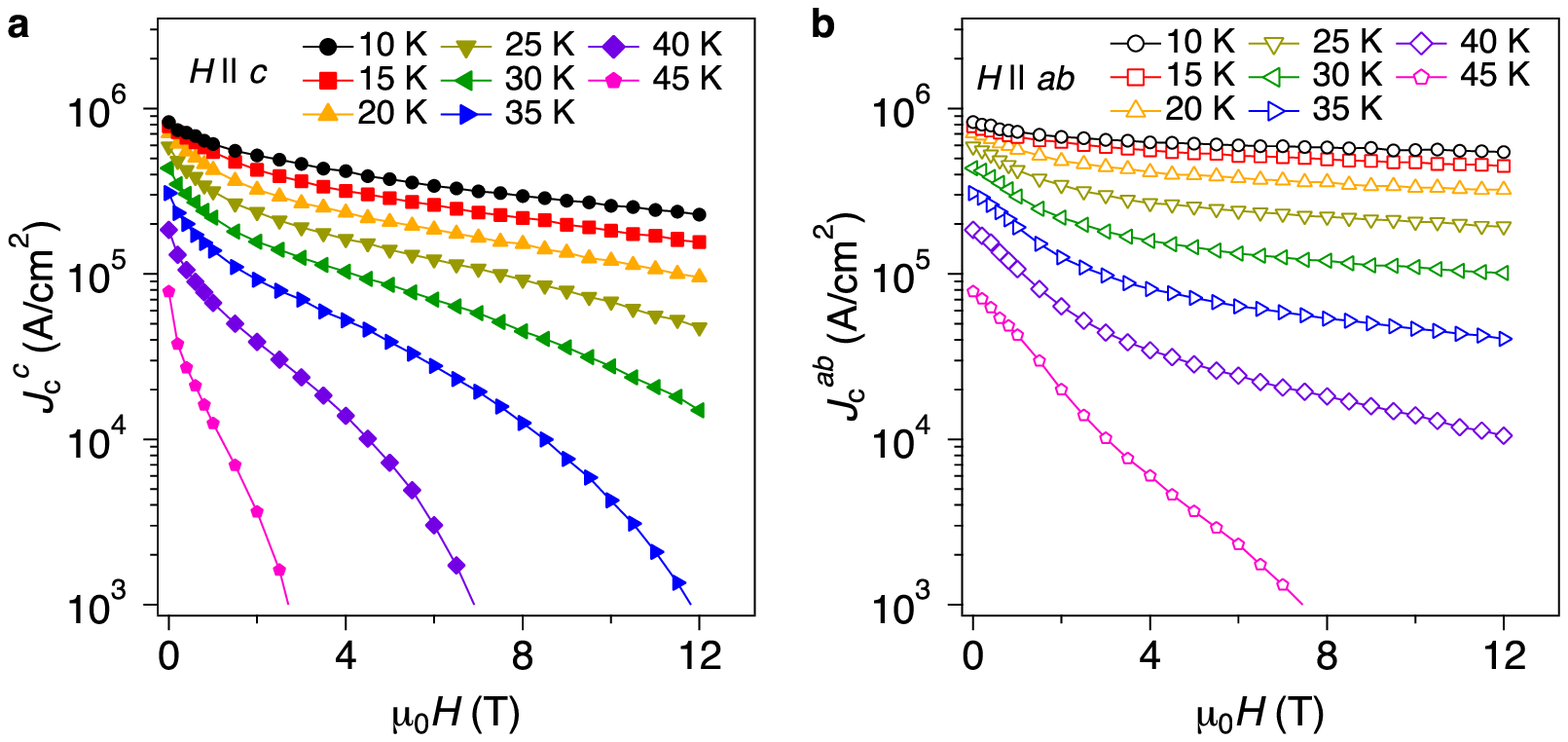}
		\caption{Magnetic field dependence of $J_{\rm c}$ measured at different temperatures for (a) $H\parallel c$ ($J^{c}_{\rm c}$) and (b) $H\parallel ab$ ($J^{ab}_{\rm c}$). For $\mu_{0}H>3$\,T, $J^{ab}_{\rm c}$ is getting more and more insensitive to $H$ with decreasing temperature.} 
\label{fig:SA2}
\end{figure}

\end{document}